\documentclass[12pt]{article}

\usepackage{bbm}
\usepackage{epsfig}
\usepackage{array}
\usepackage{float}
\usepackage{dsfont}
\usepackage{amstext}
\usepackage{rotating}
\usepackage{a4}
\usepackage{a4wide}
\usepackage{cite}

\def\ra{\rightarrow}
\def\be{\begin{equation}}
\def\ee{\end{equation}}
\def\gs{\mathrel{
   \rlap{\raise 0.511ex \hbox{$>$}}{\lower 0.511ex \hbox{$\sim$}}}}
\def\ls{\mathrel{
   \rlap{\raise 0.511ex \hbox{$<$}}{\lower 0.511ex \hbox{$\sim$}}}}
\newcommand{\obb}{0\mbox{$\nu\beta\beta$}}
\newcommand{\onbb}{neutrinoless double beta decay }
\newcommand{\ba}{\begin{array}{c}}
\newcommand{\baz}{\begin{array}{cc}}
\newcommand{\barrr}{\begin{array}{rrr}}
\newcommand{\bad}{\begin{array}{ccc}}
\newcommand{\bav}{\begin{array}{cccc}}
\newcommand{\baf}{\begin{array}{ccccc}}
\newcommand{\bea}{\begin{equation} \begin{array}{c}}
\newcommand{\eea}{ \end{array} \end{equation}}
\newcommand{\ea}{\end{array}}
\newcommand{\D}{\displaystyle}
\newcommand{\dms}{\mbox{$\Delta m^2_{\odot}$}}
\newcommand{\dma}{\mbox{$\Delta m^2_{\rm A}$}}
\newcommand{\meff}{\mbox{$\langle m \rangle$}}

\newcommand{\gsim}{\raise0.3ex\hbox{$\;>$\kern-0.75em\raise-1.1ex\hbox{$\sim\;$}}} 
\newcommand{\lsim}{\raise0.3ex\hbox{$\;<$\kern-0.75em\raise-1.1ex\hbox{$\sim\;$}}}

\begin{document}

\title{\vspace{-1cm}
\hfill {\small FERMILAB--PUB--08--099--T}\\[-0.1in]
\hfill {\small arXiv: 0804.4581 [hep-ph]} 
\vskip 0.4cm
\large \bf 
Model-Independent Analysis of Tri-bimaximal Mixing --\\
a Softly-Broken Hidden or an Accidental Symmetry?}
\author{
Carl H.~Albright$^{a,b}$\thanks{email: 
\tt albright@fnal.gov}~\mbox{ 
},~~Werner Rodejohann$^c$\thanks{email: 
\tt werner.rodejohann@mpi-hd.mpg.de}
\\\\
{\normalsize \it$^a$Department of Physics, Northern Illinois University,}\\
{\normalsize \it DeKalb, Illinois 60115, USA}\\ \\ 
{\normalsize \it$^b$Fermi National Accelerator Laboratory,}\\
{\normalsize \it Batavia, Illinois 60510, USA}\\ \\ 
{\normalsize \it$^c$Max--Planck--Institut f\"ur Kernphysik,}\\
{\normalsize \it  Postfach 103980, D--69029 Heidelberg, Germany} 
}
\date{}
\maketitle
\thispagestyle{empty}
\vspace{-0.8cm}
\begin{abstract}
\noindent  
To address the issue of whether tri-bimaximal mixing (TBM) is a 
softly-broken hidden or an 
accidental symmetry, we adopt a model-independent
analysis in which we perturb a neutrino mass matrix leading to TBM in the
most general way but leave the three texture zeros of the diagonal charged 
lepton mass matrix unperturbed.  We compare predictions for the perturbed
neutrino TBM parameters with those obtained from typical $SO(10)$ 
grand unified theories with a variety of flavor symmetries. Whereas 
$SO(10)$ GUTs almost always predict a normal mass hierarchy for the light 
neutrinos, TBM has a priori no preference for neutrino masses.  We find, 
in particular for the latter, that the value of $|U_{e3}|$ is very 
sensitive to the neutrino mass scale and ordering. 
Observation of $|U_{e3}|^2 > 0.001$ to 0.01 
within the next few years would be incompatible with 
softly-broken TBM and a normal mass
hierarchy and would suggest that the apparent TBM symmetry is an accidental
symmetry instead. No such conclusions can be drawn for the inverted and 
quasi-degenerate hierarchy spectra.

\end{abstract}

\newpage

Neutrino oscillations seem to point towards a lepton flavor 
structure completely different from that of the quark sector. 
In particular, the PMNS lepton mixing matrix has a 
very different structure 
from that of the CKM quark mixing matrix. Nevertheless, 
this seemingly 
incompatible feature can be reconciled in Grand Unified 
Theories (GUTs), 
where quarks and leptons belong to the same multiplets.  
In particular, 
GUTs based on $SO(10)$, which allow for a seesaw 
mechanism \cite{seesaw} 
without adding singlets by hand, have been 
frequently studied in the past ten 
years; see e.g.~\cite{SO10reviews}.  The models
often specify in addition a particular flavor 
symmetry with charges assigned
to the fermion and Higgs $SO(10)$ multiplets, 
although so-called ``minimal'' 
Higgs models may rely on no flavor symmetry at all.  

Another more recent approach for explaining the 
neutrino mixing scheme has
involved the introduction of lepton flavor symmetries. 
The goal of such 
models has been to reproduce the approximate 
tri-bimaximal mixing (TBM) form 
observed by Harrison, Perkins and Scott among others \cite{tri}:
\be \label{eq:Utri} 
U_{\rm PMNS} \simeq U_{\rm TBM} = \left(
\bad 
\sqrt{\frac{2}{3}} & \sqrt{\frac{1}{3}} & 0 \\
-\sqrt{\frac{1}{6}} & \sqrt{\frac{1}{3}} & -\sqrt{\frac{1}{2}} \\
-\sqrt{\frac{1}{6}} & \sqrt{\frac{1}{3}} & \sqrt{\frac{1}{2}}  
\ea 
\right) P~.
\ee
Here $P = {\rm diag}(1, \, e^{i\alpha}, \, e^{i\beta})$ 
contains the two Majorana phases, $\alpha$ and $\beta$, 
whereas the Dirac 
phase $\delta$ remains unspecified. Lepton flavor symmetries 
such as $A_4$, 
$S_4$, and $S_3$ have been applied which lead to the TBM in 
a straightforward manner \cite{A4}.  Even
more recently symmetries such as $T'$ have been introduced as an 
extended flavor symmetry in order to treat the quark sector 
as well in a self-consistent fashion \cite{T'}.

The question then arises whether the TBM symmetry is a presumably 
softly-broken hidden symmetry, or whether it is an 
accidental symmetry in nature.  Note that the 
flavor symmetries originally introduced with 
$SO(10)$ models were not designed to reproduce the 
TBM matrix per se, but 
rather were designed to reproduce quark and 
lepton mixing schemes in 
approximate agreement with the then known mixing data.  
Even with more 
refined data and fits to the data now available in the 
literature, many $SO(10)$ models have still survived.  
For reference, we quote the 
current best-fit values and 1$\sigma$ (3$\sigma$) ranges 
for the neutrino
oscillation parameters as given in \cite{data}:
\begin{eqnarray} \label{eq:data} \nonumber 
\Delta m^2_{21} & = &  7.67 \,_{-0.21}^{+0.22} \, 
\left(_{-0.61}^{+0.67}\right) \times 10^{-5}~{\rm eV}^2 \,, 
\\ \nonumber 
    \Delta m^2_{31} & = & \left\{ \baz 
-2.37 \pm 0.15 \,\left(_{-0.46}^{+0.43}\right) 
\times 10^{-3}~{\rm eV}^2  & \text{(inverted ordering)} \,, \\ 
+2.46 \pm 0.15 \,\left(_{-0.42}^{+0.47}\right) \times 
10^{-3}~{\rm eV}^2  & \text{(normal ordering)} \,,
    \ea \right. \\ 
\sin^2 \theta_{12} & = & 0.32 \pm 0.02 
\,\left(_{-0.06}^{+0.08}\right) \,,  \\ \nonumber 
    \sin^2 \theta_{23} & = & 0.45 \,_{-0.06}^{+0.09} \, 
\left(_{-0.13}^{+0.19}\right) \,,  \\ \nonumber 
 \sin^2 \theta_{13} & = & 0.0 \,^{+0.019}_{-0.000} \, 
\left(^{+0.05}_{-0.00}\right) \,.
 \end{eqnarray}

\noindent  For exact TBM, the mixing angles correspond to 
\begin{equation}
  \sin^2 \theta_{12} = \frac{1}{3}~,\quad 
\sin^2 \theta_{23} = \frac{1}{2}~,
  \quad \sin^2 \theta_{13} = 0~.
\label{eq:TBM}
\end{equation}

In order to address this issue and 
provide a partial answer to the 
question raised, we shall study 
(what we consider) reasonable deviations
from TBM and compare those results with the 
predictions of $SO(10)$ models available in the literature. 
In doing so, we will 
not assume a particular 
model leading to TBM, but instead take the 
corresponding neutrino mass 
matrix for exact TBM at face value after adopting a 
basis in which the 
charged leptons are real and diagonal, i.e., 
$U_\ell = \mathbbm{1}$. 
Let us stress here the following: we take the point of 
view that some 
unknown flavor symmetry generates TBM and work in the 
charged lepton
basis.  The flavor symmetry results obtained in any 
other basis can readily be 
cast into this form, so our results will apply in general.  
With the specified choice of basis the mass matrix, 
$m_\nu$, uniquely giving rise to TBM is 
\be \label{eq:mnutbm}
(m_\nu)_{\rm TBM} = 
U_{\rm TBM}^\ast  \, m_\nu^{\rm diag} \, U_{\rm TBM}^\dagger = 
\left(
\bad 
A & B & B \\[0.2cm]
\cdot & \frac{1}{2} (A + B + D) & \frac{1}{2} (A + B - D)\\[0.2cm]
\cdot & \cdot & \frac{1}{2} (A + B + D)
\ea 
\right)\,.
\ee
Here $m_\nu^{\rm diag} = {\rm diag}(m_1, m_2, m_3)$ 
and the parameters $A,B,D$ are in general complex and functions 
of the neutrino masses and Majorana phases:
\be \D 
A = \frac 13 \left(2 \, m_1 + m_2 \, e^{-2i\alpha} \right) ~,~~
B = \frac 13 \left(m_2 \, e^{-2i\alpha} - m_1 \right) ~,~~
\D D = m_3 \, e^{-2i\beta} ~.
\ee
In this short note we will investigate in a general manner 
deviations from the 
tri-bimaximal mixing texture. Our Ansatz is to modify 
the structure of 
the mass matrix by multiplying each element of Eq.~(\ref{eq:mnutbm}) 
with an individual complex correction factor, $\epsilon_i$: 
 \be \label{eq:mnu_dev}
m_\nu =   
\left(
\bad 
A \, (1 + \epsilon_1) & B \, (1 + \epsilon_2) 
& B \, (1 + \epsilon_3)\\[0.2cm]
\cdot & \frac{1}{2} (A + B + D) \, (1 + \epsilon_4) 
& \frac{1}{2} (A + B - D) \, (1 + \epsilon_5)\\[0.2cm]
\cdot & \cdot & \frac{1}{2} (A + B + D) 
\, (1 + \epsilon_6)
\ea 
\right)\,.
\ee
Here the complex perturbation parameters are taken to be $|\epsilon_i| 
\le 0.2$ for $i = 1 - 6$ with their phases $\phi_i$ allowed to lie 
between zero and $2\pi$. 

Note that had we chosen instead to perturb the original three parameters 
$A, \ B$ and $D$ with complex parameters 
$\epsilon_A,\ \epsilon_B$ and $\epsilon_D$, the 
neutrino masses, $m_i$ and $\Delta m^2_{ij}$ would be altered but the 
mixing matrix would remain TBM.  Instead we perturb the 
neutrino mass matrix
as above but demand that the three texture zeros in the diagonal charged 
lepton mass matrix, $m_\ell$, remain unperturbed.  The same perturbation 
prescription then applied to $m_\ell$ simply results in a diagonal phase 
transformation acting on the matrix $U_\ell$, 
which can be rotated 
away in $U_{\rm PMNS} = U^\dagger_\ell \, U_{\rm TBM}$.  

Still one may insist on applying corrections from the charged lepton
sector.  For example, with $U_{\rm PMNS} = 
U_\ell^\dagger \, U_\nu$, one can assume that $U_\nu$ corresponds to 
tri-bimaximal mixing and that the correction is given by 
\be \label{eq:Ulep}
U_\ell \simeq 
\left(
\bad
1 & \lambda & 0 \\
-\lambda & 1 & 0 \\
0 & 0 & 1
\ea
\right)~.
\ee
Then in the basis in which the charged leptons are diagonal, the 
neutrino mass matrix reads 
\be
m_\nu' = U_\ell^\dagger \, (m_\nu)_{\rm TBM} \, U_\ell^\ast ~.
\ee
In the case of a normal hierarchy where $A \simeq B \simeq \sqrt{\dms}/3 
\ll D \simeq \sqrt{\dma}$, one finds for the $e\mu$ entry that 
$(m_\nu')_{e\mu} \simeq B \, (1 - \frac DB \, \frac \lambda 2)$. 
Since $\frac DB \, \frac \lambda 2 
\simeq \frac 32 \, \sqrt{\dma/\dms} \, \lambda \ls 9 
\, \lambda$, we require $\lambda \ls 0.02$ in order to have a soft-breaking
perturbation to $(m_\nu)_{\rm TBM}$ of less than 20\%.  A similar number 
holds in the inverted hierarchy, where for 
$\alpha = \pi/2$, $A \simeq \sqrt{\dma}/3$ and 
$B \simeq -2\sqrt{\dma}/3$, 
the $\mu\mu$ entry receives the largest corrections,
roughly $8 \lambda$. Again, demanding a soft-breaking correction less than
20\% requires $\lambda \ls 0.025$. It is then easy to see from 
\cite{FPR} with $U_{\rm PMNS} = U_\ell^\dagger \, U_{\rm TBM}$ 
and $U_\ell$ from Eq.~(\ref{eq:Ulep}), the 
following small deviations from TBM are obtained:
\be
|U_{e3}|^2 = \frac{\lambda^2}{2} \mbox{ and } 
\sin^2 2 \theta_{23} \simeq 1 - \frac 14 \, \lambda^4 ~.
\ee 
The implied small values of $\lambda$ lead to $|U_{e3}|^2$ well below 
$10^{-3}$ and $\sin^2 2 \theta_{23}$ very close to 1. 
In the spirit of this note we do not tolerate a 
leptonic correction with as large a value
as $\lambda = 0.22$ for the soft breaking perturbation.  We prefer to 
work in the charged lepton basis and proceed as indicated in the previous 
paragraph.  

Radiative corrections also lead to perturbations of a 
tri-bimaximal mass matrix \cite{AKLR}. 
The relevant small parameters depend 
on the charged lepton masses, so that the $\tau$ contribution 
is enough to consider. In this limit the $e\tau$ and $\mu\tau$ 
elements are multiplied with $(1 + \epsilon_\tau)$, while 
the $\tau\tau$ entry is multiplied with $(1 + 2 \, \epsilon_\tau)$. 
The small parameter is defined as 
$\epsilon_\tau = 
c \, \frac{m_\tau^2}{16 \pi^2 \, v^2} \ln \frac{M_X}{m_Z}$, 
where $v = 174$ GeV 
and $c$ is given by 3/2 in the SM and by $-(1 + \tan^2 \beta)$ 
in the MSSM. Demanding that $|2 \, \epsilon_\tau|$ be less 
than 0.2 leads for $M_X = 10^{15}~(10^{9})$ GeV 
to $\tan \beta \ls 71.4~(97.1)$. Therefore, our corrections 
include radiative corrections up to this huge value of $\tan \beta$. 
A detailed analysis of radiative corrections to TBM can be 
found in Ref.~\cite{DGR,PR}.

Returning to the perturbed matrix in Eq. (\ref{eq:mnu_dev}), we will vary the
complex $\epsilon$ parameters, diagonalize the resulting $m_\nu$'s and 
study the predictions for the neutrino mixing angles. 
The results obtained for the perturbed mixing 
matrix will of course depend on the neutrino mass ordering and scale. 
We shall find that $|U_{e3}|$ depends most sensitively on these observables.  
Luckily, one expects a sizable improvement on its current upper limit of 
$|U_{e3}| \ls 0.2$ in the near future. We then compare our findings from 
broken TBM with predictions from successful $SO(10)$ GUTs and study how 
one may distinguish these two approaches experimentally.

Let us first consider normal mass ordering. The strategy we adopt
is as follows: we fix $m_3 = 0.050$ eV, and take as starting 
values for the other masses, $m_2 = 0.0095$ eV and $m_1 = 0.0037$ eV, 
corresponding to $\Delta m^2_{21} = 7.66 \times 10^{-5}$ eV$^2$ and 
$\Delta m^2_{31} = 2.49 \times 10^{-3}$ eV$^2$, values well within the 
center of the currently allowed region.  We shall allow for a 20\% 
variation around the initial 
values of $m_2$ and $m_1$ and vary the phases $\alpha$ and $\beta$ 
between zero and $2\pi$. Furthermore, the complex perturbation 
parameters in Eq.~(\ref{eq:mnu_dev}) are also varied within 
$|\epsilon_i| \le 0.2$ for each $i = 1 - 6$, with the full range of phases
allowed for each. For each choice of parameters the resulting mass matrix 
is diagonalized and, if the outcome is within the current $3\sigma$ 
range from Eq.~(\ref{eq:data}), the point is kept. Note that 
with a maximal perturbation of 20\% of the individual mass 
matrix elements, two entries can have a relative variation 
of 40\%, which is quite generous. 

\begin{figure}[ht]\vspace{-1cm}
\begin{center}
\epsfig{file=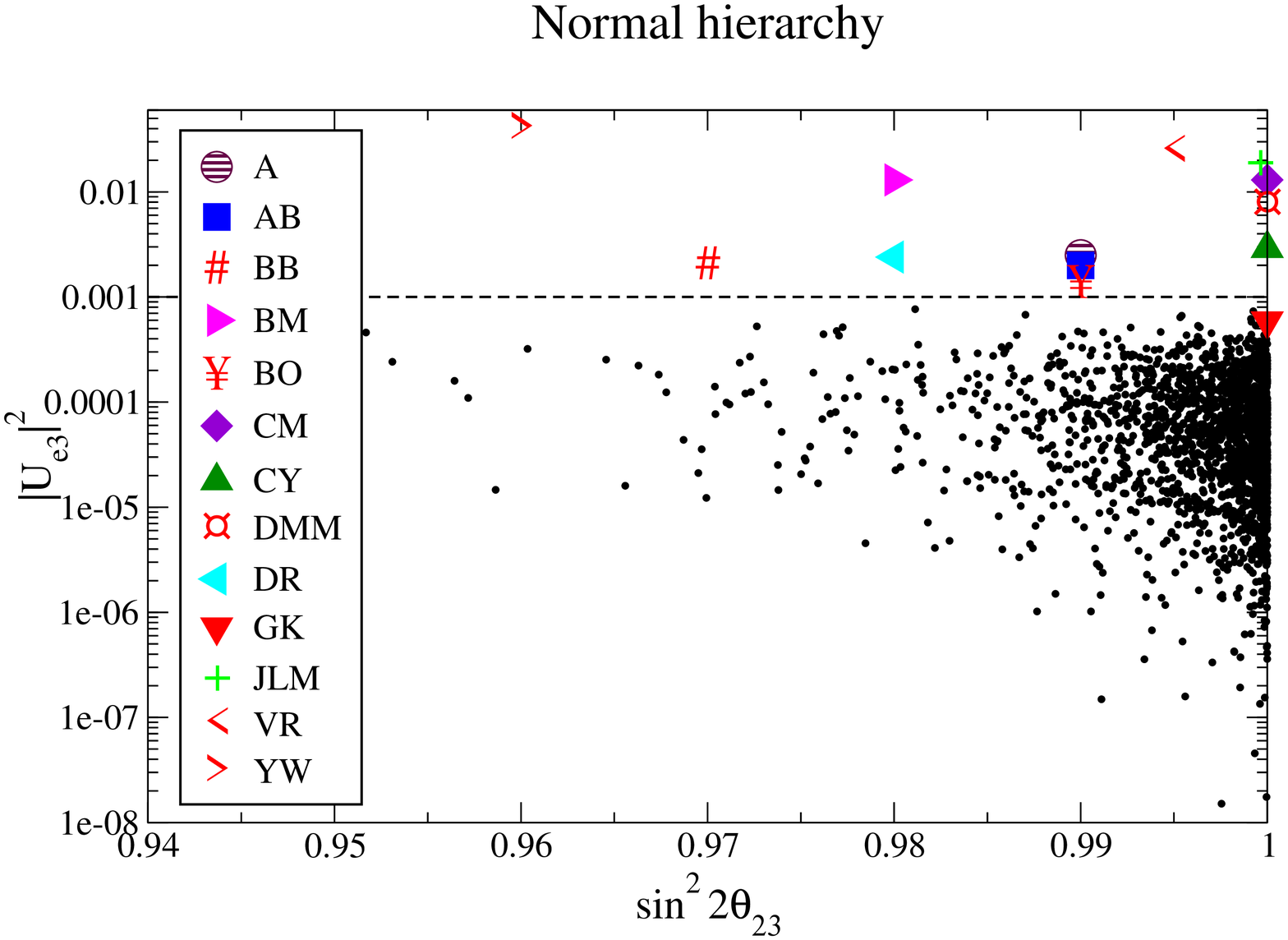,width=14cm,height=10cm} 
\end{center}\vspace{-.24cm}
\caption{\label{fig:NHa}Scatter plot of $\sin^2 2 \theta_{23}$ 
against $|U_{e3}|^2$ for perturbed tri-bimaximal mixing and a normal 
mass hierarchy. Also given are predictions of thirteen $SO(10)$ GUT models.}
%
\vspace{.3cm}
\begin{center}
\epsfig{file=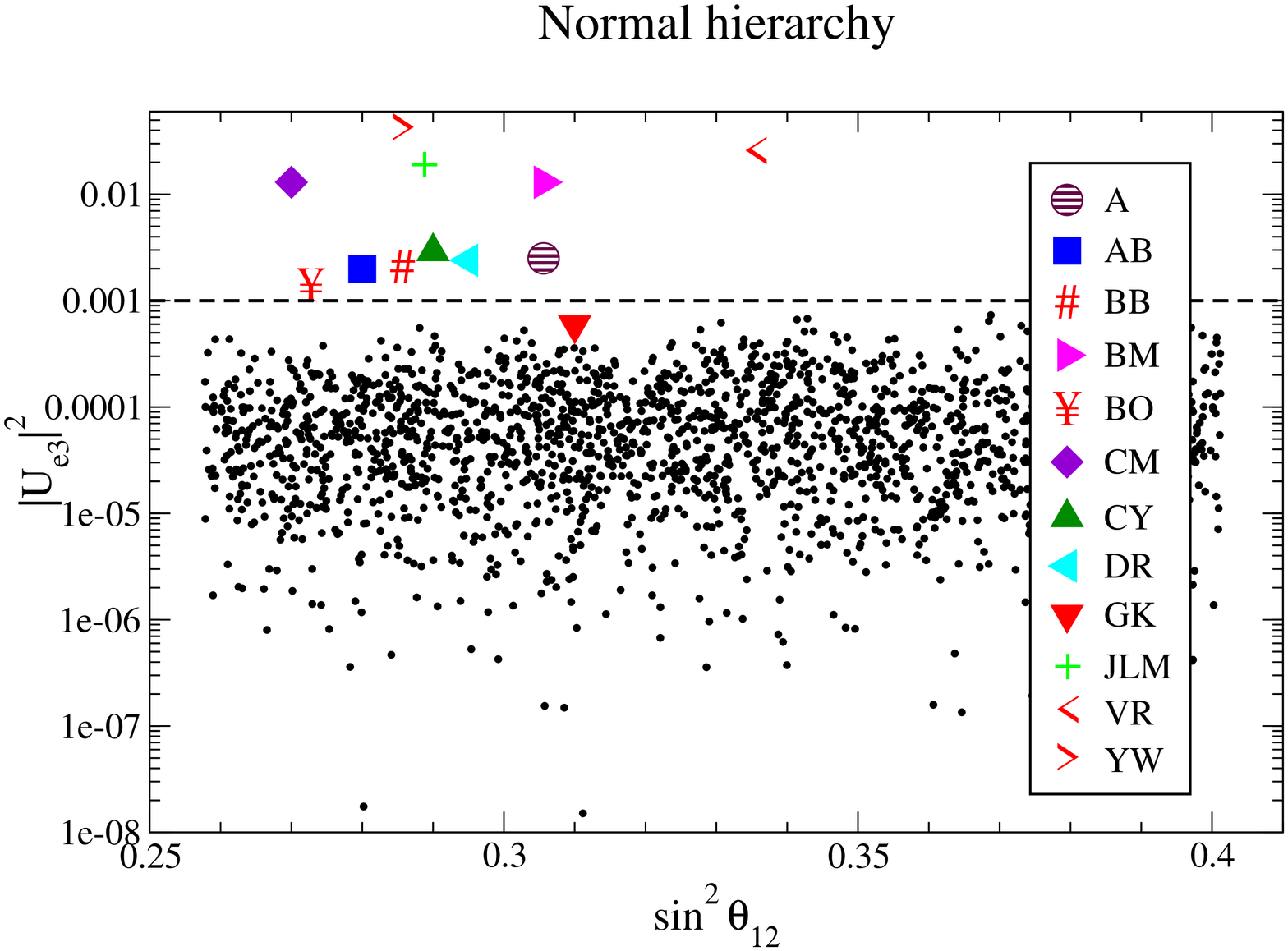,width=14cm,height=10cm} 
\end{center}\vspace{-.24cm}
\caption{\label{fig:NHb}Scatter plot of $\sin^2 \theta_{12}$ 
against $|U_{e3}|^2$ for perturbed tri-bimaximal mixing and a normal 
mass hierarchy. Also given are predictions of thirteen $SO(10)$ GUT models.}
\end{figure}

The scatter plots in Figs.~\ref{fig:NHa} and \ref{fig:NHb} 
show the results of this analysis. 
We see that $|U_{e3}|^2$ is predicted to be rather 
small and lie below $10^{-3}$. This number should be compared with the 
expected sensitivity of the Double Chooz reactor experiment \cite{DC}, 
which will start data taking in 2009, and will reach a 90\% C.L.~limit 
of 0.018 after one year with one detector, and 0.005 after 3 years of 
operation with both detectors. The Daya Bay experiment \cite{DB}, 
presumably starting after Double Chooz, is expected to improve the 
limit by another factor of two. Our results show that neither of the 
two experiments is expected to find a positive signal, if a flavor symmetry 
predicting tri-bimaximal mixing and a normal mass hierarchy is broken 
by less than 20\%. The same is true of course for the currently running 
long-baseline MINOS \cite{Minos} and OPERA \cite{OPERA} experiments. 
It will take the first generation superbeam experiments, or perhaps even 
more advanced technologies such as $\beta$-beams or neutrino 
factories, to probe $|U_{e3}|^2$ in the range below 
$\sin^2 \theta_{13} = 10^{-3}$
predicted by the perturbed TBM results.  On the other hand, 
$\sin^2 \theta_{12}$ is uniformly populated over its 
experimentally allowed range and appears to have no bearing on the issue
raised.

\begin{table}[ht]
\begin{center} 
\begin{tabular}{|c|c|c|c|c|} \hline 
Model & Hierarchy & $\sin^2 2 \theta_{23}$ & $|U_{e3}|^2$  & 
  $\sin^2 \theta_{12} $ \\ \hline \hline
A \cite{A}   & NH & 0.99 & 0.0025 & 0.31 \\ 
AB \cite{AB} & NH & 0.99 & 0.0020 & 0.28 \\  
BB \cite{BB} & NH & 0.97 & 0.0021 & 0.29 \\ 
BM \cite{BM} & NH & 0.98 & 0.013 & 0.31  \\ 
BO \cite{BO} & NH & 0.99 & 0.0014 & 0.27 \\
CM \cite{CM} & NH & 1.00 & 0.013 & 0.27 \\ 
CY \cite{CY} & NH & 1.00 & 0.0029 & 0.29 \\ 
DMM \cite{DMM} & NH & 1.00 & 0.0078 & -- \\ 
DR \cite{DR} & NH & 0.98 & 0.0024 & 0.30 \\ 
GK \cite{GK} & NH & 1.00 & 0.00059 & 0.31 \\ 
JLM \cite{JLM}& NH & 1.0 & 0.0189 & 0.29 \\ 
VR \cite{VR} & NH & 0.995 & 0.024 & 0.34 \\ 
YW \cite{YW} & NH & 0.96 & 0.04 & 0.29 \\ \hline 
S-B TBM & NH & $\gs 0.94$ & $\ls 10^{-3}$ & -- \\ 
S-B TBM & IH & $\gs 0.91$ & $\ls 10^{-2}$ & -- \\ 
S-B TBM & QD & -- & -- & -- \\ \hline
\end{tabular}
\caption{\label{tab:models}$SO(10)$ models and their 
predictions for the lepton mixing angles. If ranges are given we 
take the central value. Also given are the constraints, if any, 
on the mixing angles for the three possible mass orderings from 
the softly-broken tri-bimaximal mixing mass matrices.}
\end{center}
\end{table}

For comparison, we also give in Table \ref{tab:models} the 
results for thirteen $SO(10)$ 
GUT models, all of which involve a conventional type I seesaw mechanism 
and predict a normal mass hierarchy for the light neutrinos.  All of 
these models predict all three mixing angles 
in their currently allowed range. According to the names of the 
authors, the references are A \cite{A}, AB \cite{AB}, BB \cite{BB} 
BM \cite{BM}, BO \cite{BO}, CM \cite{CM}, CY \cite{CY}, DMM \cite{DMM}, 
DR \cite{DR}, GK \cite{GK}, JLM \cite{JLM}, VR \cite{VR} and 
YW \cite{YW}. For details, we refer to the cited works, and also 
to recent model compilations \cite{carl0,carl,rabi}.
Note that the $SO(10)$ predictions are well separated from the 
results of perturbed TBM and are accessible or more nearly accessible to 
the reactor experiments discussed above. 
The only exception is the GK model. 
We note in 
this respect that this model is very much challenged by 
its relatively large
predictions for lepton flavor violating decays like $\mu \ra e \gamma$ 
\cite{carl}. 

It is of interest to present some approximate analytical results to 
support the numerical work leading to the scatter plot in 
Fig.~\ref{fig:NHa}.  We will not try here to diagonalize the fully
perturbed mass matrix, but rather estimate the implied order of 
magnitude of $|U_{e3}|$ and $\sin^2 2 \theta_{23}$.  
A general statement, independent of the mass ordering, 
is that if the perturbation occurs only 
in the $ee$ or $\mu\tau$ entry of the tri-bimaximal $m_\nu$ 
from Eq.~(\ref{eq:mnutbm}), then the resulting mass matrix 
is still $\mu$--$\tau$ symmetric. Consequently the relation 
$|U_{e3}| = 1 - \sin^2 2 \theta_{23} = 0$ still holds in this case. 
The largest deviation of $|U_{e3}|$ from zero occurs when the 
$e\mu$ and $e\tau$ entry of the tri-bimaximal $m_\nu$ are perturbed 
\cite{PR,rabi0}. 
If we set all other perturbations to zero, the extreme case 
occurs when the $e\mu$ element is multiplied by $(1 - \epsilon)$ 
while the $e\tau$ element is multiplied by $(1 + \epsilon)$, where 
$\epsilon$ is here real. 
We can then diagonalize Eq.~(\ref{eq:mnu_dev}) and find 
(ignoring the phases and neglecting $m_1$ for simplicity) that 
\be
|U_{e3}|^2 \simeq 4 \, A^2 \, \epsilon^2  \simeq 
\frac 49 \,  R \, \epsilon^2 \ls 7 \times 10^{-4} ~.
\ee
where $R = \dms/\dma$ is the ratio of the solar and atmospheric 
mass-squared differences with 
$\dms = m_2^2 - m_1^2$ and $\dma = |m_3^2 - m_1^2|$. 
This is actually quite close to the 
numerical result, and the small discrepancy can be 
explained by the effects 
of non-zero $m_1$ and the other small terms including $\epsilon_i$. 
The smallest value for 
$\sin^2 2 \theta_{23}$ is achieved for a perturbation in the 
$\mu$--$\tau$ block of the tri-bimaximal $m_\nu$ \cite{PR,rabi0}. 
The extreme case occurs for a multiplication of the $\mu\mu$ entry 
by $(1 + \epsilon)$ and of the $\tau\tau$ entry by $(1 - \epsilon)$, 
and yields  
\be
\sin^2 2 \theta_{23} \simeq 1 - \epsilon^2 \gs 
0.96~.
\ee
This is also quite close to the numerical result.

For the inverted hierarchy case, we note that such a stable 
hierarchy is difficult to obtain in $SO(10)$ models which do not 
have a type II seesaw structure, i.e., if there is no 
direct left-handed Majorana contribution 
arising from a real or effective Higgs triplet.  
The $SO(10)$ model from Ref.~\cite{G2}, which has a negligible 
type II (triplet) contribution, is able to fit an inverted 
hierarchy, but is not very predictive in what regards the mixing 
angles. The model from Ref.~\cite{G3} concentrates on embedding 
a neutrino mass matrix with vanishing diagonal elements, 
assumes type II (triplet) dominance, and has a best-fit of 
$|U_{e3}|^2 = 0.0025$ in case of an inverted hierarchy but no other
mixing angle predictions.

\begin{figure}[ht]\vspace{-1cm}
\begin{center}
\epsfig{file=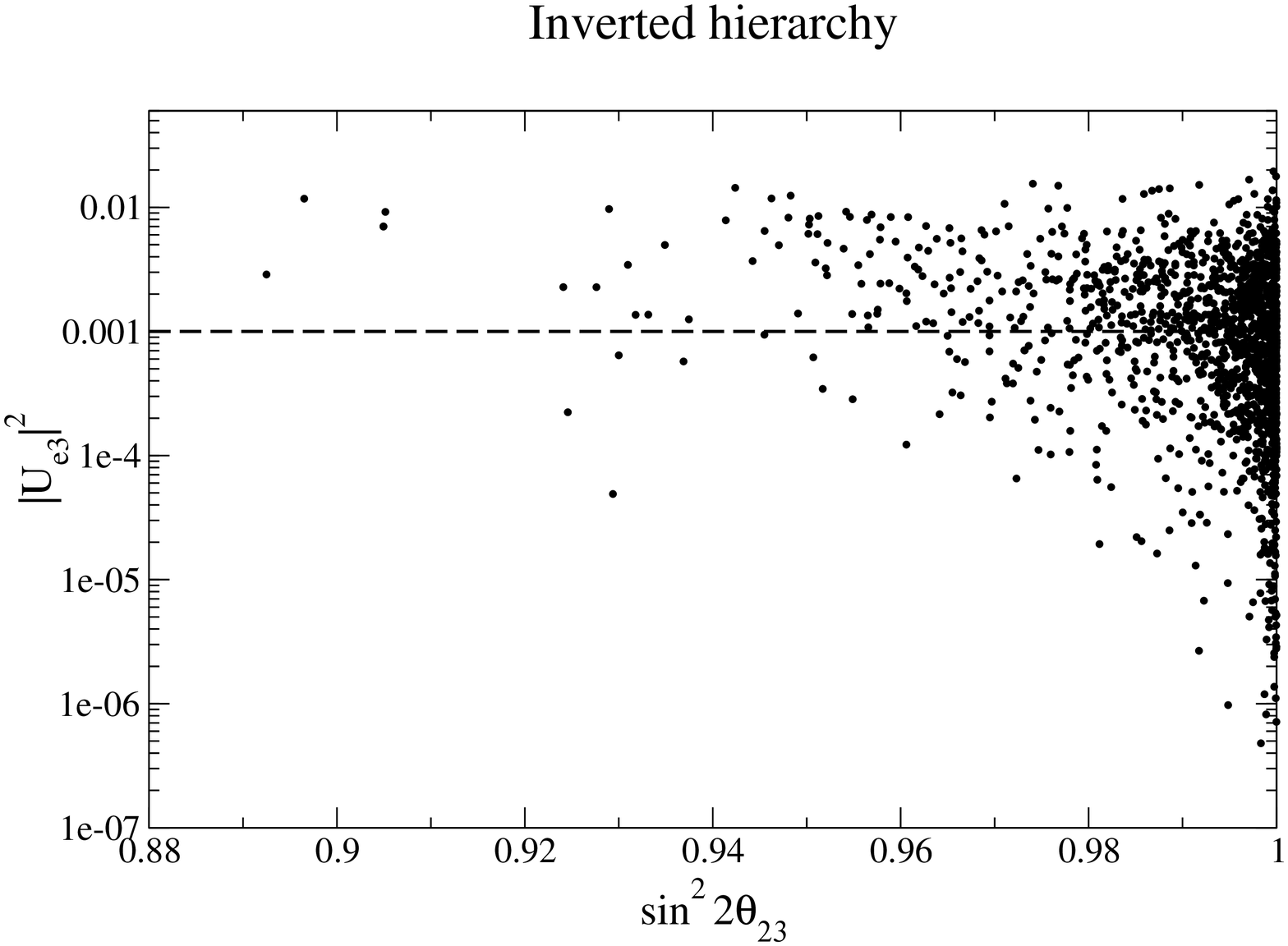,width=14cm,height=10cm} 
\end{center}\vspace{-.24cm}
\caption{\label{fig:IHa}Scatter plot of $\sin^2 2 \theta_{23}$ 
against $|U_{e3}|^2$ for perturbed tri-bimaximal mixing and an inverted  
mass hierarchy. }
\vspace{.3cm}
\begin{center}
\epsfig{file=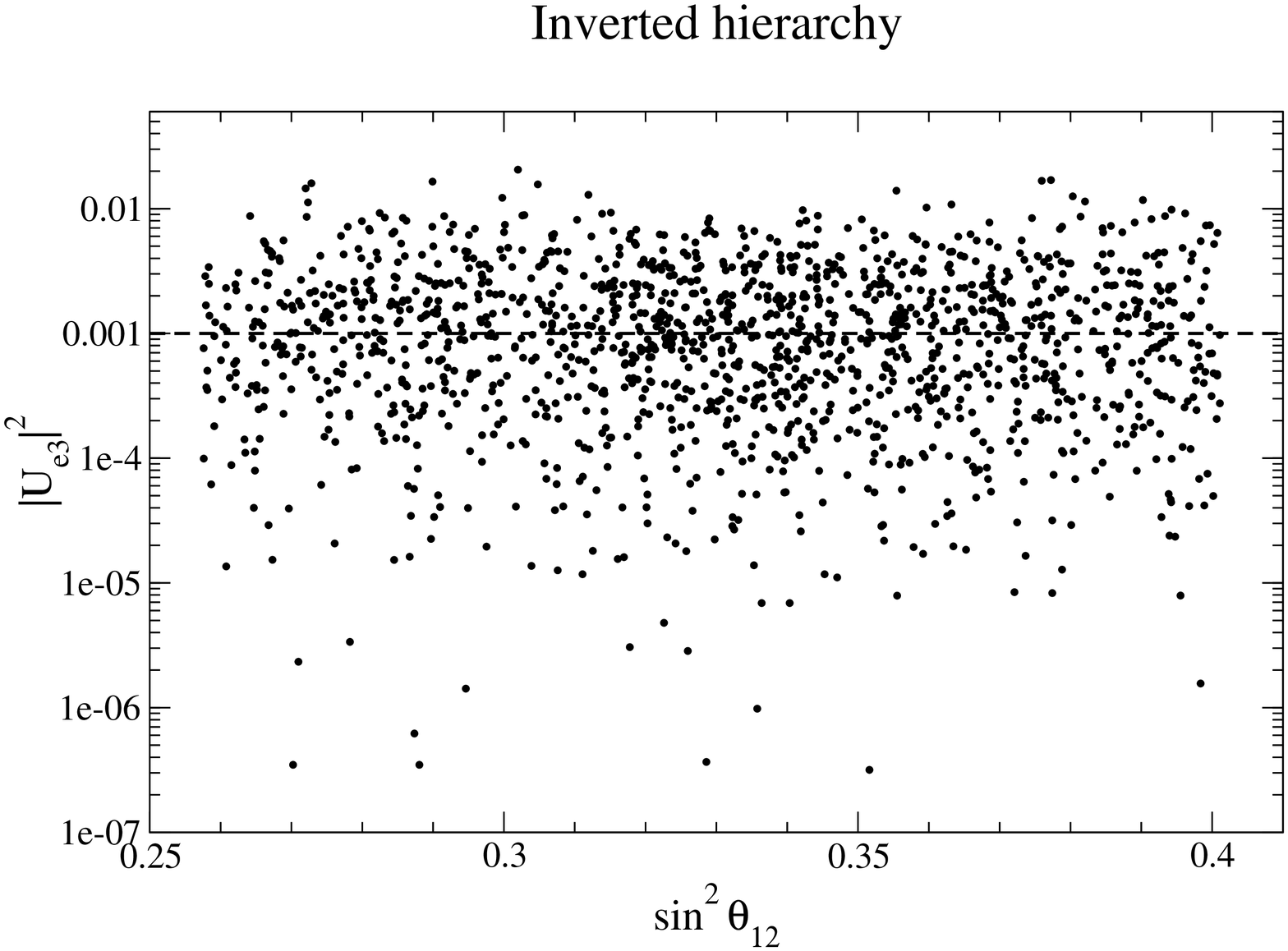,width=14cm,height=10cm} 
\end{center}\vspace{-.24cm}
\caption{\label{fig:IHb}Scatter plot of $\sin^2 \theta_{12}$ 
against $|U_{e3}|^2$ for perturbed tri-bimaximal mixing and an inverted  
mass hierarchy. }
\end{figure}

The results for perturbed tri-bimaximal matrices arising from an inverted
hierarchy are shown as scatter plots in Figs.~\ref{fig:IHa} 
and \ref{fig:IHb}.  Here we have started with 
fixed $m_2 = 0.05076$ eV, $m_1 = 0.050$ eV and $m_3 = 0.0114$ eV, 
corresponding to $\Delta m^2_{21} = 7.66 \times 10^{-5}$ eV$^2$ 
and $\Delta m^2_{31} = -2.37 \times 10^{-3}$ eV$^2$. Proceeding as in the 
normal hierarchy case we have varied the phases, $\alpha$ and $\beta$, and 
masses $m_1$ and $m_3$ within 20\% of their starting values. To guide the eye, 
we have drawn with a dashed line the value of $|U_{e3}|^2 = 0.001$, which is 
roughly the upper value found in case of a normal mass hierarchy which 
separates the perturbed results from $SO(10)$ GUT predictions. 
It is clear from the plots in Figs. 3 and 4 that $|U_{e3}|^2$ can easily 
be around, and even above, 0.01 and is therefore testable in up-coming 
reactor experiments, unlike the normal hierarchy case. Again 
$\sin^2 \theta_{12}$ is unconstrained.

\begin{figure}[ht]\vspace{-1cm}
\begin{center}
\epsfig{file=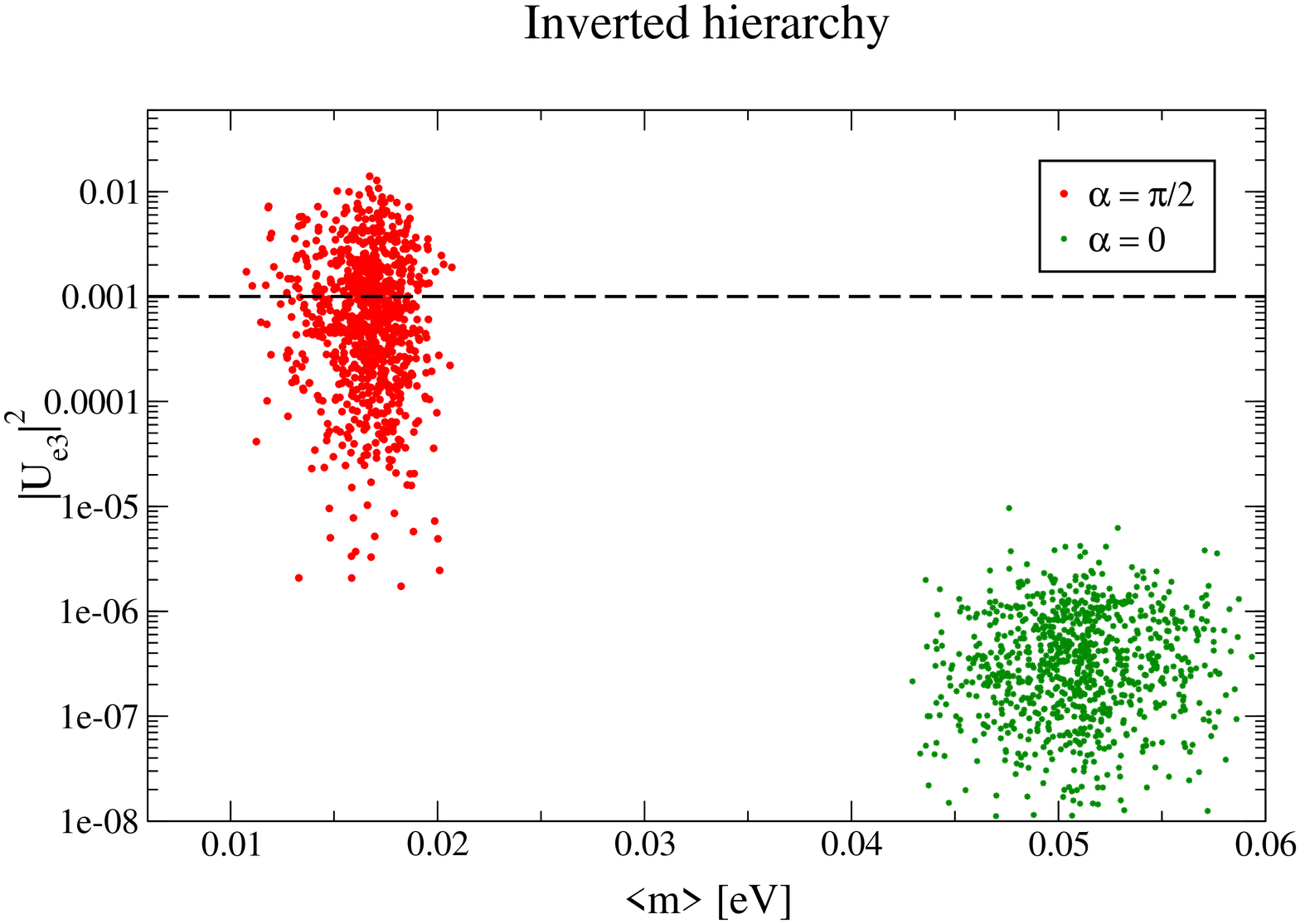,width=14cm,height=10cm} 
\end{center}\vspace{-.30cm}
\caption{\label{fig:IHphas_a}Scatter plot of the effective mass  
against $|U_{e3}|^2$ for perturbed tri-bimaximal mixing and an inverted  
mass hierarchy with two different choices of the Majorana phase 
$\alpha$, with the $\alpha = \pi/2$ cluster on the 
left and the $\alpha = 0$ cluster on the right.}
\vspace{-.1cm}
\begin{center}
\epsfig{file=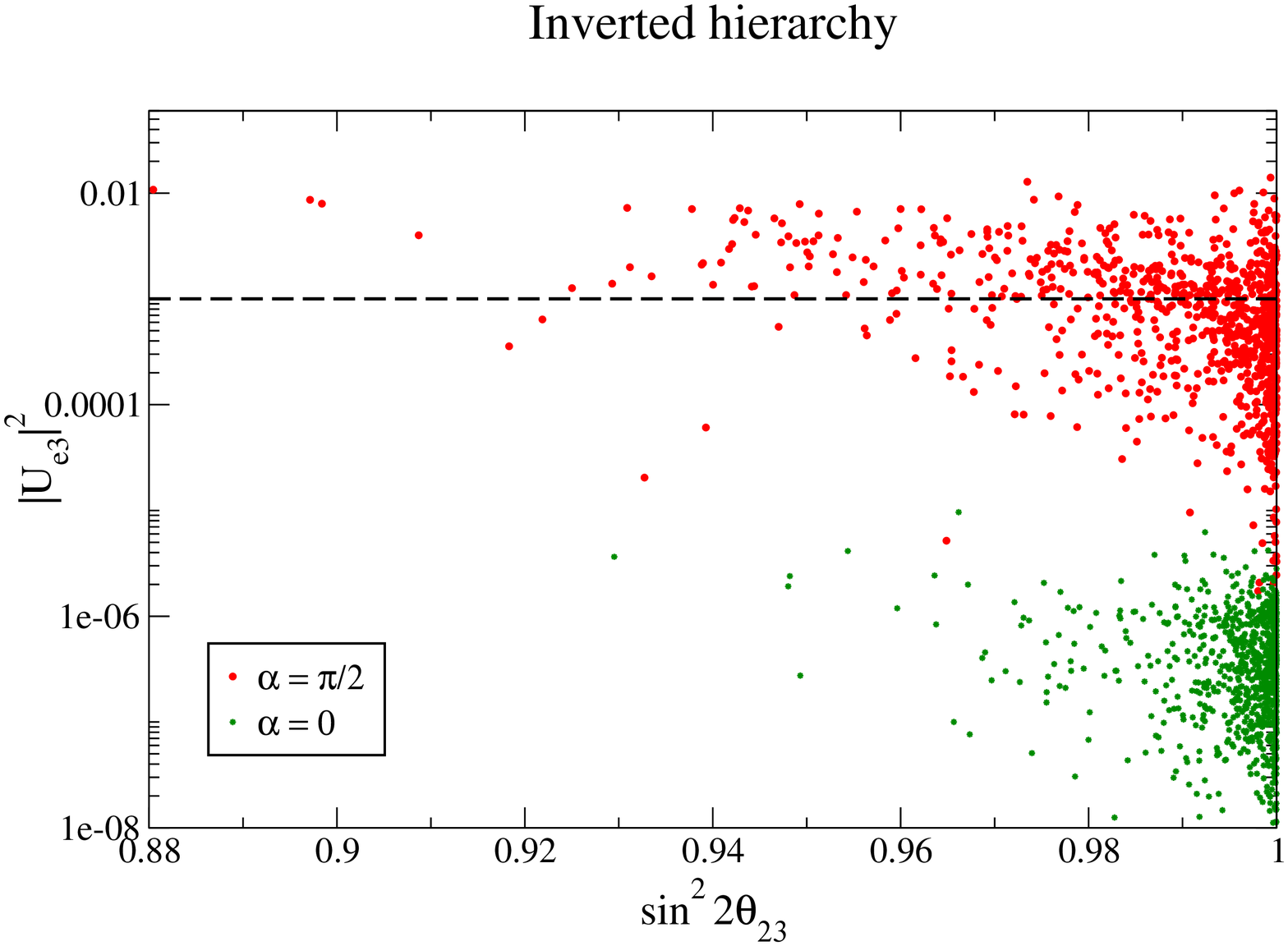,width=14cm,height=10cm} 
\end{center}\vspace{-.30cm}
\caption{\label{fig:IHphas_b}Scatter plot of $\sin^2 2\theta_{23}$ 
against $|U_{e3}|^2$ for perturbed tri-bimaximal mixing and an inverted  
mass hierarchy with two different choices of the Majorana phase~$\alpha$,
with the upper cluster referring to $\alpha = \pi/2$ and the lower to 
$\alpha = 0$.}
\end{figure}

In Figs.~\ref{fig:IHphas_a} and ~\ref{fig:IHphas_b} we show results for 
restricted values of $\alpha = 0$ and $\pi/2$ appearing in the parameters
$A$ and $B$ of the TBM neutrino mass matrix.  We find that the largest 
values of $|U_{e3}|^2$ typically occur if the phase $\alpha$ is around 
$\pi/2$. This value implies that the effective mass governing neutrinoless 
double beta decay (\obb) takes its minimally allowed value: 
$\meff \simeq \sqrt{\dma} \, \cos 2 \theta_{12} 
\simeq \frac 13 \, \sqrt{\dma}$.  For the normal hierarchy case, $\meff$
is expected to be smaller still.  On the other hand, if the phase $\alpha$
is zero or $\pi$, $\meff \simeq \sqrt{\dma}$, and $|U_{e3}|^2$ is tiny.  
We also note that $\theta_{23}$ can deviate more sizably from maximal 
mixing, if neutrinos are inversely ordered, and that the largest 
deviation occurs for $\alpha = \pi/2$ as shown in Fig.~\ref{fig:IHphas_b}.

Turning to analytic estimates for inverted ordering, 
for $\alpha = \pi/2$ 
one has $A \simeq \sqrt{\dma}/3$ and $B \simeq -2\sqrt{\dma}/3$. Consider 
first a perturbation of 
the $e\mu$ entry with $(1 + \epsilon)$ and of the $e\tau$ entry with 
$(1 - \epsilon)$ for real $\epsilon$. In this case 
\be
|U_{e3}|^2 \simeq  \epsilon^2 \left( \frac{8}{81} 
+ \frac{16}{27} \, \frac{m_3}{\sqrt{\dma}}
\right) \ls 10^{-2} 
\mbox{ and } 
\sin^2 2 \theta_{23} \simeq 1 - \left(\frac{16}{9}\right)^2 \, 
\epsilon^2 \gs 0.87~.
\ee
In the case of $\alpha = 0$, we have $B/A \simeq \frac 16 \, \dms/\dma$ 
and find that $|U_{e3}|^2$ is at most 
of order $(\epsilon \, B/A)^2 \simeq 10^{-6}$ 
and therefore completely negligible. In the case of perturbed 
$\mu\mu$ and $\tau\tau$ entries, 
$\sin^2 2 \theta_{23} \simeq 1 - \epsilon^2 \gs 0.96$. The agreement 
with the figures is quite reasonable.

\begin{figure}[ht]\vspace{-1cm}
\begin{center}
\epsfig{file=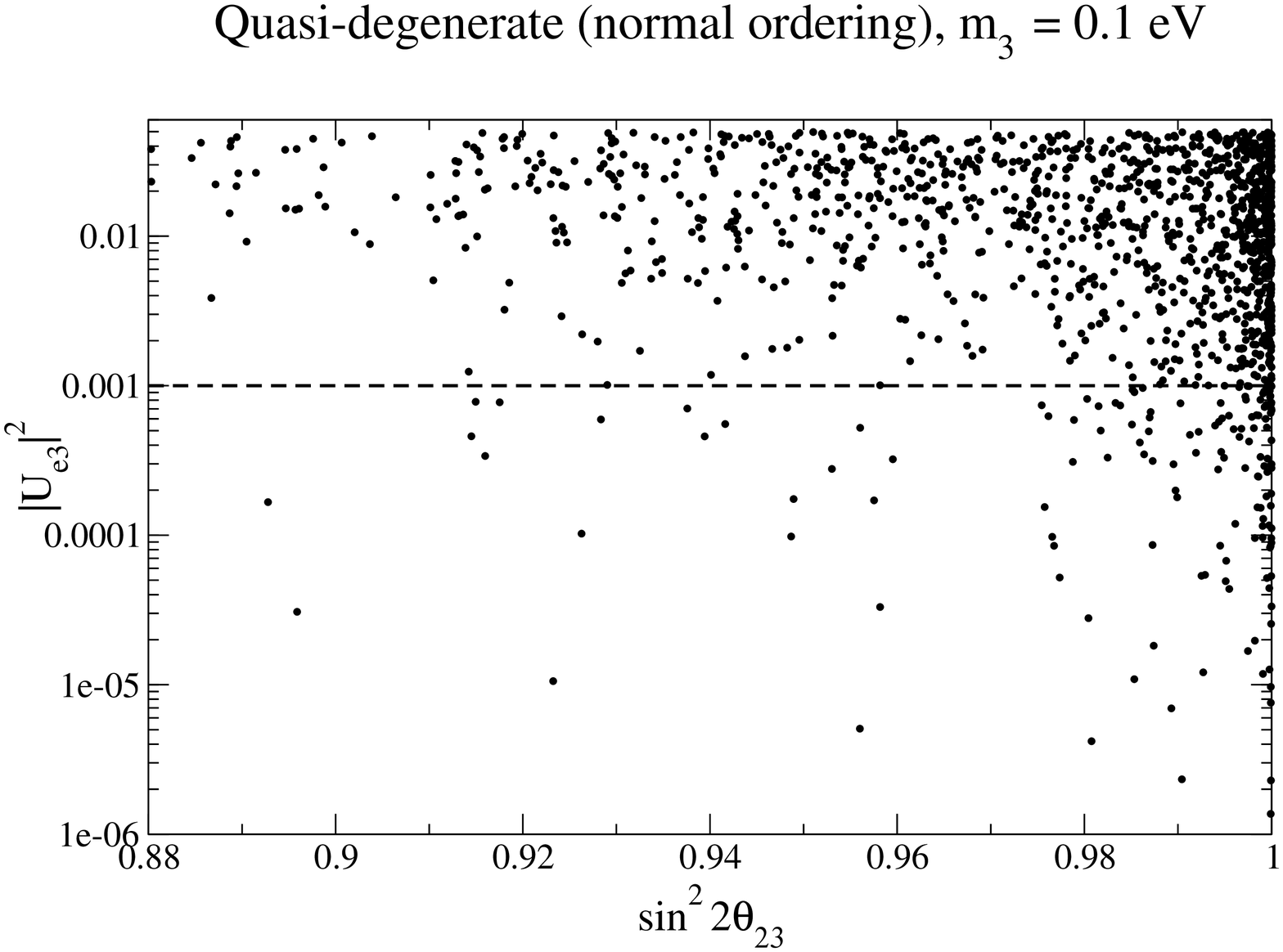,width=14cm,height=10cm} 
\end{center}\vspace{-.24cm}
\caption{\label{fig:QDa}Scatter plot of $\sin^2 2 \theta_{23}$ 
against $|U_{e3}|^2$ for perturbed tri-bimaximal mixing and 
quasi-degenerate neutrinos. }
\vspace{.3cm}
%
\begin{center}
\epsfig{file=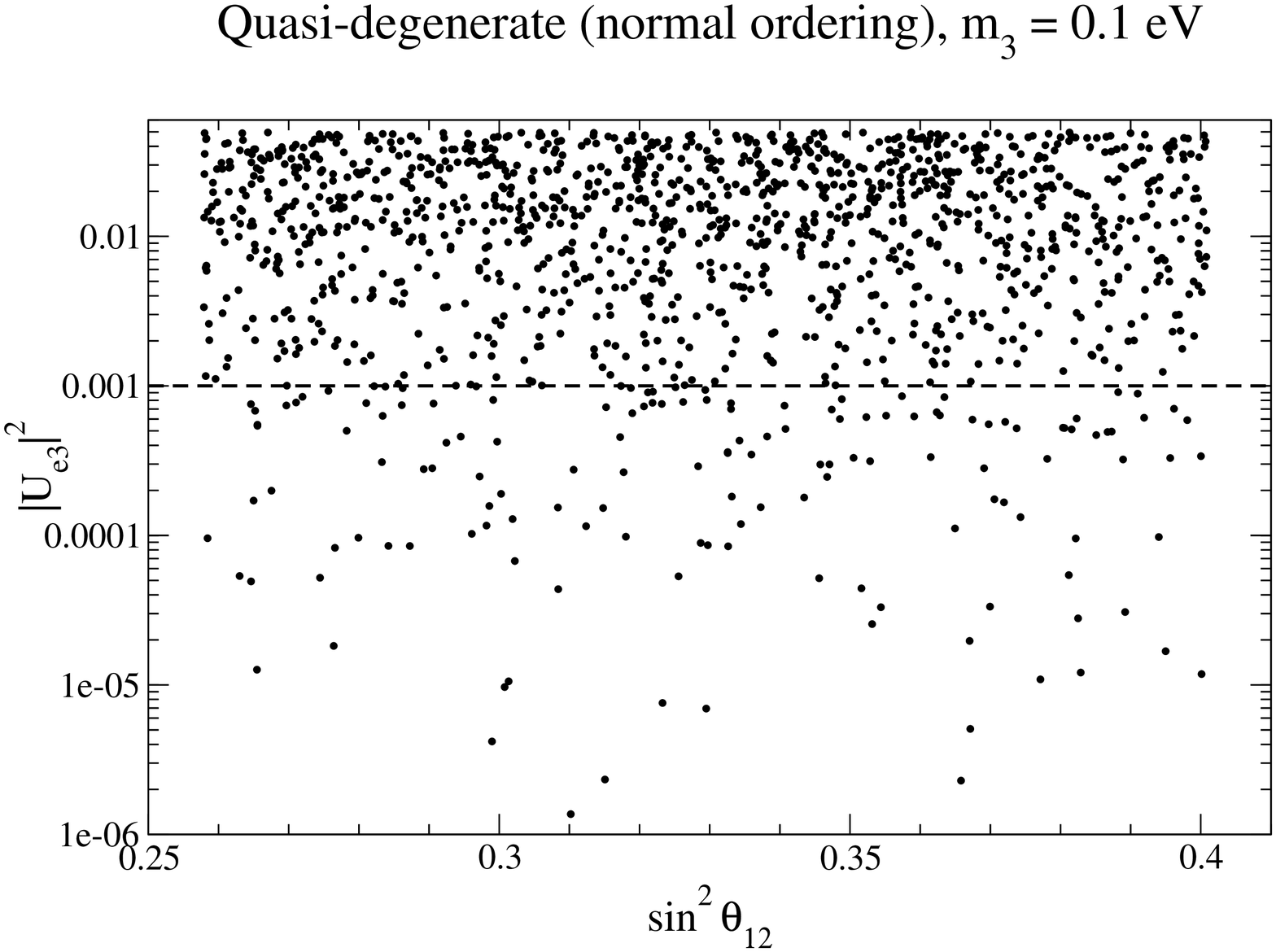,width=14cm,height=10cm} 
\end{center}\vspace{-.24cm}
\caption{\label{fig:QDb}Scatter plot of $\sin^2 \theta_{12}$ 
against $|U_{e3}|^2$ for perturbed tri-bimaximal mixing and 
quasi-degenerate neutrinos. }
\end{figure}

For completeness we study also the case of quasi-degenerate neutrinos. 
For definiteness, we consider the case of normally ordered neutrinos, 
choosing  fixed $m_3 = 0.1$ eV, $m_2 = 0.08778$ eV and 
$m_1 = 0.08735$ eV, corresponding to 
$\Delta m^2_{21} = 7.59 \times 10^{-5}$ eV$^2$ 
and $\Delta m^2_{31} = 2.37 \times 10^{-3}$ eV$^2$. The procedure is the
same as before and the results are shown in Figs.~\ref{fig:QDa} and 
\ref{fig:QDb}. Plots for an inverted ordering of quasi-degenerate 
neutrinos look basically identical. 
As expected, both $\sin^2 2 \theta_{23}$ and $|U_{e3}|^2$ 
deviate more sizably than before from their initial, tri-bimaximal 
values.  Since this quasi-degenerate case is more similar to the inverted
one, it is not surprising that the interplay of the 
Majorana phase $\alpha$, 
the effective mass and the deviation from maximal $\theta_{23}$ 
and zero $\theta_{13}$ are similar.  Analytically, one finds that an 
enhancement of roughly $m_1^2/\dma \simeq 4$ occurs for the upper limits
of $|U_{e3}|^2$.  All mixing angles are populated in 
their allowed ranges.

In summary, we have raised the question whether the approximately 
tri-bimaximal mixing observed in the lepton sector results from a hidden 
symmetry or whether it is accidental.  Early proposed mass matrix models 
based on $SO(10)$ family symmetry were not designed to lead to 
tri-bimaximal mixing, although a number of them are still successful.  
To study this issue we have adopted a model-independent approach and 
perturbed the TBM neutrino mass matrix elements about their central 
values by 20\% in the lepton flavor basis.  The charged lepton mass 
matrix is trivially perturbed by keeping the off-diagonal three texture 
zeros intact.  We found that $|U_{e3}|^2$ and the neutrino mass scale 
and ordering are of importance in the problem. In general the value of 
$\sin^2 \theta_{12}$ is not constrained, and precision measurements of 
its value will not help in settling the issue raised. The other mixing 
parameter, $\sin^2 2 \theta_{23}$ has only limited impact.  A most 
striking result obtained is that for a normal neutrino mass hierarchy, 
the predicted perturbed TBM values of $|U_{e3}|^2$ lie below $10^{-3}$,
while the type I seesaw $SO(10)$ models typically predict 
values above this.  For an inverted or quasi-degenerate neutrino mass 
hierarchy, on the other hand, only two of the studied $SO(10)$ models 
apply, while the allowed perturbed values of $|U_{e3}|^2$ can range 
noticeably higher, up to the present experimental limit.  An interesting 
correlation with the value of the effective mass governing \onbb is 
observed. While the question posed is unanswerable at this time, we 
can conclude that observation of $|U_{e3}|$ within the next few years 
would be incompatible with TBM and a normal mass hierarchy.  Clearly 
it will be necessary to determine both $|U_{e3}|$ and the mass 
hierarchy in order to address the posed question.

\vspace{0.3cm}
\begin{center}
{\bf Acknowledgments}
\end{center}

One of the authors (C.H.A.) thanks Manfred Lindner and members of the 
Max-Planck-Institut f\"{u}r Kernphysik for their kind hospitality while
the research reported here was carried out.  The work of 
W.R.~was supported in part by the Deutsche Forschungsgemeinschaft 
in the Transregio 27 ``Neutrinos and beyond -- weakly interacting 
particles in physics, astrophysics and cosmology''.

\newpage
\pagestyle{empty}

\end{document}